# The Human-GenAI Value Loop in Human-Centered Innovation: Beyond the Magical Narrative

**Camille Grange**
Department of Information Technology
HEC Montreal, Montreal, Canada
camille.grange@hec.ca

**Théophile Demazure**
Department of Information Technology
HEC Montreal, Montreal, Canada
theophile.demazure@hec.ca

**Mickaël Ringeval**
Department of Information Technology
HEC Montreal, Montreal, Canada
mickael.ringeval@hec.ca

**Simon Boudreau**
Department of Analytics, Operations and Information Technology
UQAM, Montreal, Canada
bourdeau.simon.2@uqam.ca

**Cédric Martineau**
Clarta Conseil
cedric@clarta.ca

## Abstract

Organizations across various industries are still exploring the potential of Generative Artificial Intelligence (GenAI) to automate a variety of knowledge work processes, including managing innovation. While innovation is often viewed as a product of individual creativity, it more commonly unfolds through a collaborative process where creativity intertwines with knowledge. However, the extent and effectiveness of GenAI in supporting this process remain open questions. Our study investigates this issue using a collaborative practice research approach focused on three GenAI-enabled innovation projects conducted within different organizations. We explored how, why, and when GenAI could effectively be integrated into design sprints—a highly structured, collaborative process enabling human-centered innovation. Our research identified challenges and opportunities in synchronizing AI capabilities with human intelligence and creativity. To translate these insights into practical strategies, we propose four recommendations for organizations eager to leverage GenAI to both streamline and bring more value to their innovation processes: (1) establish a collaborative intelligence value loop with GenAI; (2) build trust in GenAI, (3) develop robust data collection and curation workflows, and (4) embrace a craftsman's discipline.

## 1. Introduction

The use of computing technologies to support or automate work has long captivated academics and practitioners (Zuboff, 1988), and this interest has reached yet another peak since the advent of generative artificial intelligence (GenAI). GenAI, recently popularized with ChatGPT and similar tools, is a subset of artificial intelligence (AI) that is deemed to excel in common knowledge-based tasks—like summarization, transcription, clarification, categorization and content generation[1].

GenAI is expected to be a transformative driver for knowledge work, which involves the cognitive processing of information to generate value-added outputs (Alavi & Westerman, 2023). The emerging empirical evidence and market data support this forecast. For example, an experiment showed that when they used ChatGPT, the time needed by midlevel professionals to complete a writing task diminished by 40% and the output quality rose by 18% (Noy & Zhang, 2023). Another study conducted in a more complex knowledge work scenario showed a comparable outcome, with gains exceeding 40% in output quality and a 25% improvement in time efficiency (Dell'Acqua et al., 2023). Industry analysis from McKinsey also suggests that GenAI could lead to substantial financial benefits, potentially generating up to $4.4 trillion annually across various sectors like customer operations, marketing, sales, software engineering, and research and development (Chui et al., 2023).

Beyond the immediate appeal of using GenAI for productivity gains at the individual task level, the McKinsey report also mentions that "as generative AI continues to develop and mature, it has the potential to open wholly new frontiers in creativity and innovation" (Chui et al. 2023, p.11). In the present study, we explore this uncharted area, which is crucial and intriguing for practitioners for two primary reasons.

First, innovation is the "lifeblood (...) for every organization trying to be successful today" (Bancel et al., 2022, p.3). Despite its critical importance (reflected in the adage "innovate or die"), innovation remains a challenging endeavor (Pisano, 2015) that involves the resolution of complex business problems (Kinni, 2017). Recent research by the Boston Consulting Group reported that 83% of companies saw innovation as a top three priority, but that only 3% were ready to translate their priorities into results, revealing a surprising paradox: "Companies have never placed a higher priority on innovation—yet they have never been as unready to deliver on their innovation aspirations." (Manly et al., 2024, p.3) Given the transformative effects of GenAI on knowledge work—an essential ingredient of innovation—its potential to improve readiness by facilitating the process of innovation is substantial.

---

[1] Generative AI uses machine learning models to recognize patterns in vast datasets, enabling it to create new content based on user inputs, referred to as prompts. The nature of inputs and outputs may vary (e.g., text, images, audio, video) depending on the tool.

Second, while innovations thrive on teamwork (Edmondson, 2012), the prevailing discourse in academia and industry focuses on how GenAI affects individual knowledge work. For example, Alavi and Westerman (2023) suggest that GenAI can help reduce workers' cognitive load by automating structured tasks, boosting their cognitive capacity for unstructured tasks, and improving their learning process. Jarrahi (2018) explains why and how AI can augment human decision-making rather than replace human decision-making. However, as Benbya et al. (2024) highlighted, the dynamics of human-machine *teams*—including how roles and tasks are distributed between humans and GenAI—remain a crucial yet underexplored issue. The field of information systems (IS) has a longstanding history of exploring the intersections of digital technology and team dynamics (Bostrom et al., 1993; DeSanctis & Gallupe, 1987; Poole & DeSanctis, 1990), aligning with ongoing practical interests in this area (Leonardi, 2021). This all points to the relevance of a deeper examination of how a rapidly evolving technology like GenAI can be effectively embedded into collaborative work settings to foster collective creativity and drive innovation.

In this article, we explore the roles of GenAI within innovation teams by detailing a year-long journey through GenAI-enhanced projects at three organizations. This investigation provides an opportunity to reassess the traditional notion that innovation primarily revolves around individual creativity and idea generation (Gilson & Litchfield, 2017; Gotlar & Hutley, 2024). Echoing earlier critics such as Harvey (2014), we argue that this perspective is limited in its ability to capture the complex and collaborative nature of innovation. Instead, aligning with Design Thinking principles, we approach innovation as a structured, collaborative problem-solving process (Brown, 2008; Liedtka, 2018). This perspective allows us to explore the varied applications of GenAI across different cognitive tasks—involving both divergent and convergent thinking—and different stages in the innovation process—from framing the problem to prototyping and testing solutions.

Our study was designed to document the variations in GenAI application and to stimulate discussions on how to optimize its integration into team creative processes. This investigation uncovered significant synergies between human teams and AI capabilities, particularly emphasizing the crucial role of human input in divergent thinking tasks such as problem exploration and solution ideation. In these stages, it is vital for teams to deeply engage with and fully "own" both the problems they explore and the solutions they devise. Conversely, GenAI showed its distinct value in convergent tasks such as aligning on a problem and deciding which solutions to test. Its ability to clarify, synthesize, and organize human thoughts can significantly reduce cognitive load for team members. These insights carry significant implications for innovation practitioners, suggesting that strategic integration of GenAI can not only enhance teams' ability to effectively use their domain knowledge but also optimize their efforts and time management. Beyond a mindful distribution of roles, our research underscores the importance of carefully managing human teams to foster trust in human-AI collaboration, diligently building robust data workflows, and adopting a craftsman-like approach to practice.

Our paper begins with an overview of the latest research insights at the intersection of artificial intelligence, innovation, and creativity. We then share findings from our collaborative practice research, emphasizing the evolving application of GenAI across three distinct innovation projects. We discuss the challenges encountered during these projects along with the key lessons that emerged. Following this, we provide practical recommendations for effectively integrating GenAI into innovation practices. The paper concludes by exploring strategies for overcoming practical challenges and briefly addressing the implications of our findings for future research.

## 2. When Generative AI Meets Creativity and Innovation

The exploration of GenAI's potential in enhancing creative and innovative processes in organizations marks a pivotal shift in the exploration of artificial intelligence's capabilities. This expansion is particularly significant considering that creativity has long been regarded as a distinct hallmark of humanity (Csikszentmihalyi, 1996) and that people remain particularly averse to AI-generated creative content (Köbis & Mossink, 2021). While creativity and innovation are interrelated, they remain distinct concepts. Creativity, broadly defined as the ability to conceive something original and useful (Runco & Jaeger, 2012), is often considered essential for innovation, which encapsulates the application of creative ideas to transform products, services, business models, or strategies, thus addressing complex challenges to enhance the status quo (Osterwalder & Pigneur, 2010). This section outlines the challenges and opportunities presented by GenAI in transforming creativity and innovation, focusing on its role in (1) co-creative endeavors and (2) human-centered innovation processes.

### 2.1 Humans and GenAI as Partners in Co-Creation

Whether humans or AI is more creative is still up for debate and empirical investigation. For instance, while Joosten et al.'s (2024) study reported that AI-generated ideas were superior in terms of novelty and customer value compared to ideas developed by humans alone, Doshi and Hauser (2023) observed that ideas generated by GenAI tended to be more homogeneous than those created by humans, raising a flag on GenAI over-reliance (Q. Liu et al., 2024). These limitations have led to an emphasis on the value of a careful and sustainable integration of AI in a way that supports rather than replaces or risks diminishing human creativity, ensuring that GenAI enhances instead of dominates the creative process. The synergy between humans and GenAI, often described in terms such as co-design (Vartiainen et al., 2025), human-AI co-creativity (Rafner et al., 2023) or human-AI symbiosis (Jarrahi, 2018; Jarrahi et al., 2023), has thus emerged as a promising frontier for exploring the intersection of GenAI, creativity, and innovation.

Pre-GenAI research demonstrated the potential of AI to act as an effective partner that amplifies the creative capabilities of human teams without overshadowing their essential roles. For instance, it has been shown that AI-driven chatbots can be instrumental in maintaining the flow and focus of brainstorming sessions in human teams (Debowski et al., 2022), mitigating the pressure experienced by anxious team members during idea generation (Hwang & Won, 2021), and managing discussions in design tasks aimed to capture deeper insights into user needs (Bittner & Shoury, 2019). With a shift in focus from traditional text-based AI bots to more advanced GenAI models, prompting has emerged as a critical component in human-AI co-creation, especially relevant in creative domains where prompts are used to generate visual representations from textual inputs (V. Liu & Chilton, 2022; Oppenlaender, 2022). This transition has sparked a dynamic area of research centered on developing methods for effectively interacting with a model to elicit the specific knowledge necessary for task completion (Noack et al., 2025; Robertson et al., 2024).

Research on the synergies between humans and GenAI in co-creative work is still in its infancy, and the contrasting findings in emerging works highlight the nuanced nature of humans x GenAI collaboration. On the one hand, for example, Eapen et al. (2023) identify several benefits of humans and AI joining forces: GenAI can stimulate divergent thinking, enabling individuals and teams to forge innovative concepts by linking disparate ideas in novel ways. Additionally, it can refine ideas by merging and enhancing various concepts to create stronger propositions and mitigate human expertise bias by introducing unconventional ideas that expand the realm of possibilities. In the same vein, Memmert and Bittner's (2024) study suggests that joint (human+GenAI) work can foster flexibility in ideation (i.e., breadth of topic exploration), an important performance metric in brainstorming. On the other hand, La Scala et al. (2025) observed that ideas generated with GenAI assistance were not more novel than those conceived by humans alone. More concerningly, Gotlar and Hutley (2024) reported that while GenAI-assisted idea generation helped avoid poor ideas, it also resulted in more mediocre ideas, suggesting that individuals may become overconfident in their problem-solving abilities when using GenAI. Moreover, downstream dangers have been noted, such as GenAI overshadowing unique human input and producing novel ideas that non-experts may find difficult to translate into feasible and actionable solutions (Rafner et al., 2023).

More broadly speaking, the interest in humans co-creating with increasingly agentic and multimodal forms of AI is sparking new inquiries into more fundamental questions, such as, who's in charge of creative problem solving (Dwivedi & Banerjee, 2024), how can the key experiential aspects of the creative process like intentionality and self-expression be preserved (Dahlstedt, 2021), how to leverage the flexibility of multimodal interactions in innovation practices (Peng et al., 2024), and how to develop synergies that enhance the capabilities of human actors and enable new collaboration patterns (Schwabe et al., 2025).

### 2.2 Integrating GenAI within Human-Centered Innovation

While innovation is often linked to individual creativity and divergent thinking, in practice, it unfolds through a complex interplay of creativity and cognitive effort that necessitates both divergent (exploring various options) and convergent (selecting and operationalizing options) thinking (Medeiros et al., 2023). To fully understand the capabilities of GenAI for innovation (Steiger et al., 2024; Sundberg & Holmström, 2024), it is thus crucial to move beyond the concept of GenAI as merely a collaborator on isolated tasks like ideation. This shift leads to two critical considerations.

First, effective innovation processes demand a deep understanding of user needs, anchored in empathetic design approaches typical of human-centered innovation (Brown, 2008). This approach encompasses various collaborative tasks—defining the problem, generating ideas, and concluding with idea evaluation and solution validation—which are interdependent and likely to vary in the extent to which automation is feasible or desirable (Marrone et al., 2024). On that matter, Cropley et al. (2023, p. 32) argued that "No matter how good AI technology becomes, the reason why we are creative (problem definition and solution validation) remains the job of humans," suggesting a potentially more nuanced role and value for GenAI within a user-centered innovation process than conveyed in more optimistic views (Holmström & Carroll, 2024).

Second, when technology is integrated into this framework, the innovation process becomes a socio-technical system where creativity emerges from the synergistic interactions between people (team members), structure (workflows), tasks (design activities), and technology (Ciriello et al., 2024). Understanding and optimizing these interactions poses a significant challenge but is crucial in leveraging GenAI's potential to enhance team creativity and, ultimately, to devise solutions that respond effectively to complex challenges.

In summary, research on the use of GenAI for enhancing creative problem-solving is dynamic and signals both promising outcomes and unresolved challenges. The mixed results so far stem largely from the nascent and rapidly evolving nature of GenAI technologies, along with the varied ways individuals and teams appropriate these tools. This critical juncture calls for a rigorous examination of GenAI's role within collaborative, human-centered innovation processes to understand better how, when, and why GenAI can be integrated into empathy-driven innovation frameworks to augment, rather than overshadow, the unique and invaluable human elements of creativity and innovation.

# 3. Insights From a Learning Journey into GenAI-Driven Innovation

## 3.1 Research Methodology

We explored the integration of GenAI within three design sprints conducted in different organizations from April to November 2023. Design sprints are rooted in design thinking, a paradigm for innovating through a user-centric approach (Martin, 2009; Verganti et al., 2021; Wangsa et al., 2022). They involve a small, multidisciplinary team that works intensively over limited time to propose an innovative solution to a complex problem (Zeratsky, 2016). The detailed sequence of activities within the original version of design sprints developed by Knapp et al. (2016) is presented in Appendix A. Our focus on design sprints in this research was influenced by their prominence in digital innovation (Dell'Era et al., 2020) and their condensed, time-boxed nature. This structure made them well-suited to our research objective as each task within a sprint (e.g., framing the problem, ideating, prototyping) plays a well-defined role in the problem-solving process. By examining how GenAI was utilized across this process, we were able to gain a detailed and dynamic understanding of its practical integration and impact.

We adopted a collaborative practice research methodology (Mathiassen, 2002), featuring a close collaboration between a practitioner and four academic researchers. The practitioner, an early adopter of GenAI in design sprints, was central to our study. His first responsibility was to facilitate the design sprints. Facilitators organize, guide, document, and manage the team's process, ensuring clarity of goals and efficient time use. They lead discussions, encourage participation, and handle conflicts to maintain focus on the sprint's objectives. They also aid in synthesizing ideas and feedback, especially during the prototyping and testing phases. In the studied projects, the facilitator chose and manipulated GenAI tools (primarily ChatGPT) in the preparation and managing of the team's activities. Additionally, the practitioner documented the activities conducted within the sprints, their timing, the application of GenAI (prompts, outputs), participants' behaviors, and the achieved outcomes. This documentation, which included extensive visual boards used to organize and lead the team's activities, was shared with the academic collaborators for discussion and analysis.

The academic researchers did not directly influence the implementation of GenAI during the sprints but contributed significantly to three main areas. First, they collected additional data by conducting semi-structured interviews with five individuals after their participation in the GenAI-augmented sprints. These interviews, which lasted about an hour each, were designed to gather deeper insights into human reactions to GenAI within the innovation process and triangulate data to enhance the study's robustness. Second, they enriched discussions by introducing theoretical and empirical insights from the recent literature. Third, they led the analysis phase, which relied on interpretive principles (Klein & Myers, 1999). In short, these principles invite a dynamic process of comprehension, where the researchers constantly move from a preliminary understanding of the parts (e.g., the use of GenAI within each sprint) to a holistic view of the whole (e.g., the means and value of combining GenAI and human work in a design sprint) and back to the parts again, thereby refining the understanding of each component within the broader context. Appendix B offers a detailed synthesis of how interpretive research principles were applied in our study.

Inspired by these principles, we (practice and academic researchers) collectively undertook both within-project and cross-project analyses, structured around twelve sensemaking sessions detailed in Appendix C. During these sessions, we focused on five levels of analysis: (1) *roles distribution*—we investigated how the problem solving process, encompassing individual and team-based activities with and without GenAI, was allocated within each project; (2) *value, challenges, and pivots*—we examined the benefits derived from using GenAI, the obstacles encountered, as well as the pivotal moments in the facilitator's practice; (3) *reasons and mechanisms*—we reflected on why and how various instances and patterns of human-GenAI collaboration created value; (4) *reflections and lessons*—we abstracted the emerging knowledge to distill context-specific learnings from each project; and (5) *synthesis*—we took stock of emerging, context-specific knowledge to subsequently develop broadly applicable, actionable guidelines for innovation practitioners. These activities were interconnected, and we continually moved between them to ensure that our recommendations accurately reflected the lessons learned from each project. Appendix D visually represents the analysis process that led to our insights and practical recommendations.

### 3.2 The Design Sprint Framework Used in the Projects

The same design sprint framework was applied across the three projects, facilitating comparisons in the use of GenAI throughout the process. The framework slightly deviates from Jake Knapp's original model at Google Ventures (detailed in Appendix A) by allocating more time to the problem space. This modification responds to the enduring challenge of effectively scoping the problem in an innovation project (Root-Bernstein, 2003), aligning our approach more closely with the principles of design sprint 3.0 (Vetan, 2018).

The sprints, structured into six main phases split between a problem and a solution space, were spread over a few weeks to fit the schedules of the participating organizations. The **problem space** is dedicated to exploring and defining the sprint's challenge, while the **solution space** focuses on imagining, prioritizing, and validating a solution. Design sprints are underpinned by two key meta-cognitive processes: *divergent thinking*, which involves generating and making sense of various ideas and possibilities, and *convergent thinking*, which is about aligning on actionable priorities and solutions. The overall process is thus often depicted through a "double diamond" diagram (Figure 1), illustrating repeated episodes of divergence and convergence in both spaces.

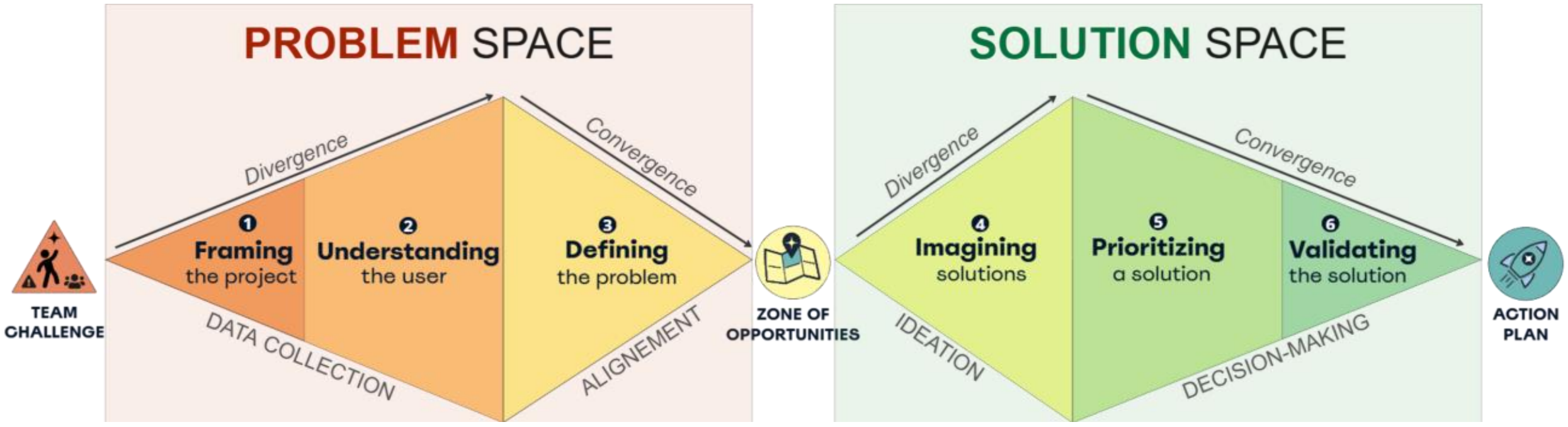

Figure 1. Overarching Design Sprint Framework Used in the Three Innovation Projects

In the three projects, the goal in the *problem space* was to define a zone of opportunities by articulating a precise problem statement ("how might we meet [business goal] through satisfying [user need]"). This involved three stages.

**(1)** Framing the project is about exploring the issue and identifying the most significant complex challenge related to the organization's business objective. It includes identifying business goals, key issues, and innovation opportunities, while understanding the overall context.

**(2)** Understanding the user focuses on understanding user needs and frustrations in the experience journey surrounding the prioritized challenge. This phase focuses on empathy and gaining a deep understanding of users so that the designed product or service truly meets their needs.

**(3)** Defining the problem involves reframing the challenge to be solved into a problem statement (broad enough to encourage creativity yet focused enough to be actionable) and setting success criteria to align the team on the opportunity area for innovation and guide the rest of the process.

The goal of the *solution* space was to craft and test a prototyped solution to inform future actions; this was achieved in three steps:

**(4)** Imagining solutions aims to encourage creativity and the exploration of a wide range of perspectives to propose ideas for innovative solutions. The goal is to broaden the realm of possibilities to discover new and potentially disruptive approaches to solving the problem.

**(5)** In the prioritizing phase, team members discuss the generated ideas and select the most promising solution concept based on impact potential, feasibility, and alignment with user needs. They refine this concept into more tangible forms by creating a user test flow (a brief user experience often depicted in the form of a text-based scenario with basic visuals), designing a visual storyboard, and formulating interview questions to validate the key elements of the concept, such as those that are new or risky.

**(6)** Validating the solution involves developing a tangible prototype (e.g., cartoon, mobile interface wireframe, website mock-up – they can be digital or physical and vary between low and higher fidelity) for the solution and assessing its desirability through user testing.

### 3.3 Application and Highlights from the Projects

In this section, we delve into the use of GenAI in the three studied innovation projects, emphasizing the key activities performed as well as critical lessons and turning points in the practice. Figure 2 illustrates the primary characteristics of these projects.

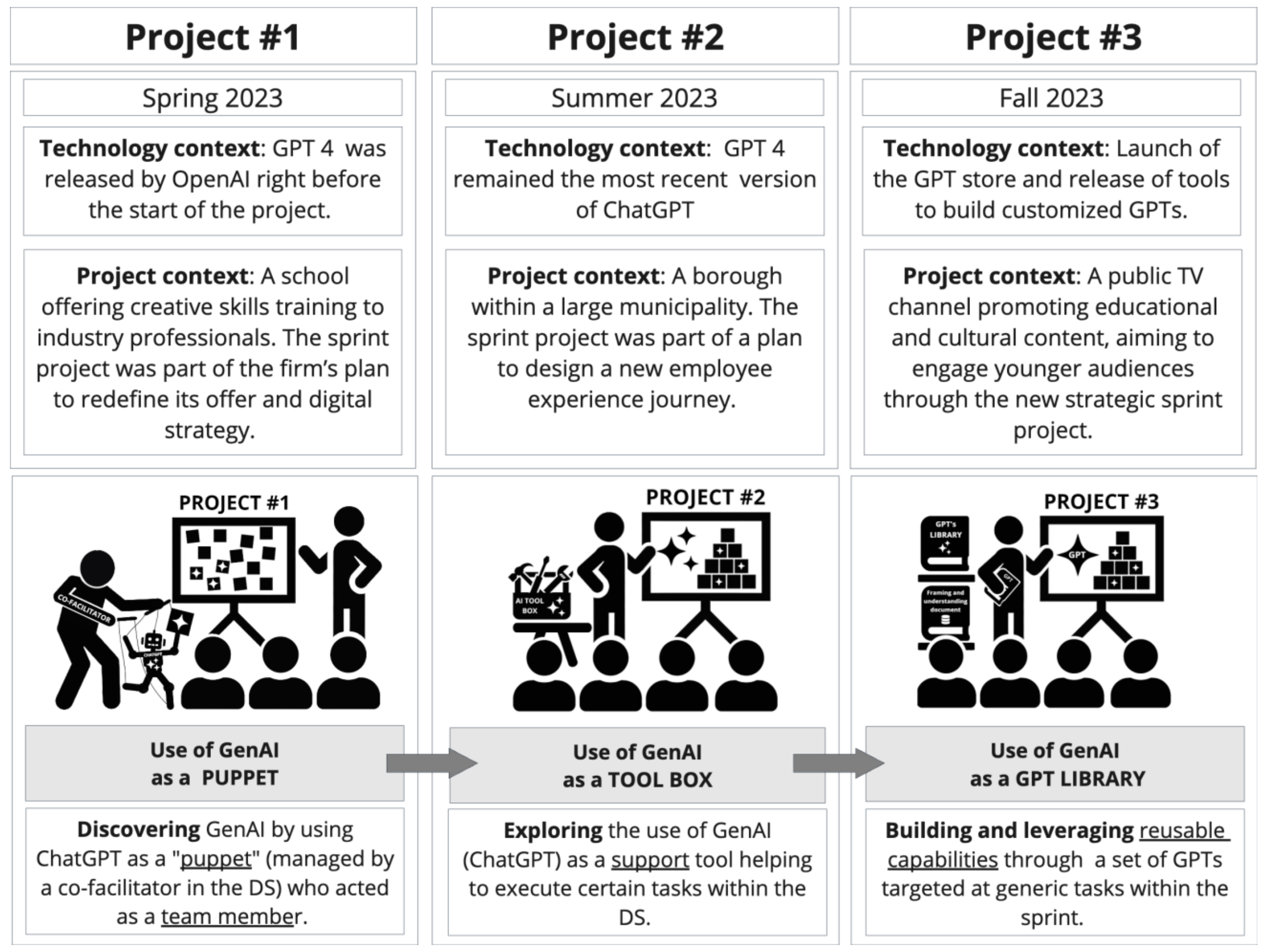

Figure 2. Overview of the three Design Sprint (DS) Projects

**Project #1: Discovering ChatGPT's Capabilities**

The first project took place at a creativity school engaged in democratizing the creative process for industry professionals. The aim was to find an innovative way to better showcase the school's educational programs. The sprint team included seven human participants, one fewer than usual, and introduced ChatGPT as an additional participant named J.A.K.E.—short for Jovial, Analytical, Knowledgeable, Eloquent—inspired by Jake Knapp, the creator of the DS methodology. The role of J.A.K.E. was that of a guest contributor rather than a decision-maker, and it was excluded from tasks that required deep contextual understanding. A human co-facilitator prompted ChatGPT, ensuring that J.A.K.E.'s contributions aligned with the project's main objectives.

In this project, the initial week was dedicated to **project framing.** This started with key stakeholder consultations led by the facilitator to draft an initial project *brief* and was followed by a *three-hour workshop* with the entire team to refine and prioritize project needs and challenges. During this session, the human participants in the team were introduced to J.A.K.E. Like its human counterparts, J.A.K.E. identified and presented what it considered the most significant and complex challenge, drawing on textual excerpts from the project *brief* provided by the co-facilitator. The excerpts used to prompt J.A.K.E. served as a foundational *project document* occasionally reused for subsequent exercises. Interestingly, the issue J.A.K.E. highlighted coincided with the main concern identified by the rest of the team.

In the **understanding** phase, a *three-hour workshop* included collaborative exercises designed to build a unified understanding of the target users, context, and desired outcomes. Each participant, including J.A.K.E., actively contributed by individually generating ideas for value propositions and problem statements. The workshop concluded with a human vote on the most promising problem to further specify. Interestingly, although the final decision was made by human team members, J.A.K.E.'s input, which suggested that "the problem is that our product offering is complicated to understand," closely mirrored the team's selected focus: "We need to be careful about the lack of clarity with our offer." Human participants, independently of J.A.K.E., also proposed three success criteria for the sprint.

During the second week, a *three-hour* ***alignment*** *workshop* was held to synchronize the team's understanding of the problem and targeted users, and to specify the opportunity area and success factors moving on to the solution space. J.A.K.E.'s proposed problem statement garnered 75% of the team's votes, making it the most popular. The team then developed multiple "how might we...?" questions to examine in mode depth the opportunities associated with the selected problem. The facilitator had intended to use J.A.K.E. to generate some of these questions, but limited time, data management challenges, and inadequate prompts restricted this effort.

In the *three-hour* ***ideating*** *workshop*, J.A.K.E. kick-started the session by presenting baseline ideas tailored to the chosen opportunity area, prompted by the co-facilitator using the *project document* updated with data from the previous workshop. This initiation guided the human participants, who also generated their own ideas, toward more innovative solutions. The workshop also involved transforming the generated ideas from short texts to more elaborate three-step sketches.

Before the two subsequent *three-hour **prioritizing** workshops*, the facilitator prepped by using ChatGPT and experimenting with DALL-E to format J.A.K.E.'s proposed concept on a digital post-it with text and drawings, similar to human participants' submissions. The AI's output was noticeably artificial but included for consideration. During the workshops, the team voted on and narrowed down the top concepts, eventually selecting one to develop further. J.A.K.E. was less involved due to the co-facilitator's limited time to integrate relevant descriptions for top concepts. Between sessions, the co-facilitator refined the prompts, enabling J.A.K.E. to contribute more effectively to the final activities (crafting a test scenario and formulating hypotheses to be tested).

During the **validation** phase, the facilitator independently used ChatGPT to generate text for the prototype and its supporting materials. After completing this, he carried out user testing, employing ChatGPT once more to assist in synthesizing the interview transcripts.

**Learnings from Project #1 and Pivot -** The client organization viewed this initial experiment of integrating GenAI into the innovation process as successful, particularly praising J.A.K.E. for its neutrality and ability to generate relevant and concise information as well as provide inspiration.

> "*J.A.K.E. really knocked it out of the park with how it perfectly pulled together all that we were trying to say, making sure we don't forget anything and helping us cope with information overload" (Project 1, participant #1) ... "J.A.K.E does not have the emotional baggage we have, it is not influenced by local politics, and it contributes to variety, which we know is a key ingredient to creativity*" (Project #1, Team member #2)

While the DS offered clear benefits, it also presented significant challenges. Managing the vast amount of textual data generated proved logistically cumbersome, especially as the co-facilitator was interacting with ChatGPT in real-time during the workshops. This process highlighted the need for better data organization. Additionally, the necessity for a co-facilitator appeared impractical, and the effectiveness of using GenAI as an anthropomorphized participant was questionable.

> "*ChatGPT's great strength is synthesis. No matter how much data you give it, it synthesizes beautifully using the keywords you provide. However, handling all that data in real-time was a headache, and honestly, trying to make the AI seem more human just felt weird to me. It's like we're trying to force it to be something it's not, and I'm not sure we're getting the best out of it that way."* (Project #1, Facilitator)

It was thus decided that the use of GenAI would evolve in two main ways in the subsequent project. First, instead of replacing a team member, ChatGPT would strictly be used as a supportive tool for the facilitator. This shift was driven by three lessons learned in Project #1: (1) the impracticality of requiring a co-facilitator to manage ChatGPT during the sprint, (2) the need to maximize contextual input by adding another human participant to feed the AI, and (3) the realization that the enthusiasm design sprints generate within organizations makes it hard to justify replacing a human team member with GenAI. Secondly, the use of GenAI would need to become more structured. The facilitator envisioned developing a repertoire of prompts, organized by sprint phase, for easier access and reuse. He would also focus on improving data quality by creating specialized prompts to process, clean, and format text data into a structured output. GenAI would thus be used both to support the collaborative activities per se and to structure the data so that it would be efficiently used as input.

**Project #2: Exploring ChatGPT as a Support Tool**

In the second project, the DS was held at a local municipality to develop an innovative onboarding experience for new employees. The lessons learned in Project #1 (particularly the need to improve data quality and structure) led to a significant evolution in how work was conducted within the problem space. While **project framing** and **user understanding** still involved a *brief* and extensive consulting of the organizational and project context, the two workshops were eliminated. Instead, a highly focused ten-question survey was sent to team members, followed by an individual meeting with each of them to delve deeper into their responses, which were consolidated with the help of ChatGPT. These meetings helped to refine the creation of a focal persona and contribute to a *framing and understanding document,* a structured text containing all synthesized information generated about the sprint's challenge. Additionally, they served to introduce the use of ChatGPT in the sprint through practical examples, address any concerns, and foster acceptance among the team. The subsequent *alignment workshop* (**defining**) was maintained to engage team members (only human participants in this project) in activities that helped them assimilate the collective results and agree on a shared perspective on the specific problem to address.

In the ***imagining*** *workshop*, the facilitator used ChatGPT to propose a set of initial ideas and to help collective sensemaking by rearticulating human-generated ideas that lacked precision and classifying them. This could be done more effectively thanks to the use of the *pre-developed prompt repertoire* together with the *framing and understanding document*.

Before the two subsequent *three-hour **prioritizing** workshops*, the facilitator used ChatGPT to synthesize textual descriptions of each participant's proposed concept, gathering relevant inputs for the following team exercises. Due to concerns about time constraints and complexity, the facilitator opted not to introduce an additional AI-generated concept. In the first workshops, the team voted for a winning concept and proposed testing hypotheses (without any GenAI assistance). In the second workshop, after some background work by the facilitator to translate these into text, a set of prepared prompts helped produce a user test flow. Both human participants and GenAI generated detailed scripts, including questions aligned with previously established sprint criteria and questions for assessing the prototype's effectiveness. It was remarkable here that seven out of eight participants voted in favor of the GenAI-generated artifacts. Similar to project #1, GenAI was used in the **validation** phase to facilitate the labor-intensive tasks of generating text for the prototype, polishing user testing questions, and synthesizing interview transcripts.

**Learnings from Project #2 and Pivot –** The use of ChatGPT in this second project confirmed some of the advantages discovered by the facilitator and participants from the first project, such as information consolidation and inspiration capabilities. Additionally, the value-cocreation process became more evident, with GenAI playing a key role in clarifying and refining team members' thoughts, while team members also actively revised GenAI outputs as needed.

> "*ChatGPT just nailed it every time. It helped us to phrase our thinking adequately… what we wanted to say. It was concise and often ended up being the go-to choice. Plus, the ability to tweak and fine-tune things as we went was reassuring. We knew right from the start not to just take it at face value. Sometimes, it would focus onto details that were insignificant to us (…) Sometimes we needed less general synthesis and more personalized input"* (Project 2, Team member #1)

> "*I believe that information from a person is invaluable. Adding one more person in this sprint brought new ideas and information and contributed to the synergy for the project. Therefore, it was worth adding someone to have more diversity around the table. (...) ChatGPT's great strength is synthesis and standardizing the quality of human participants' inputs. No matter how much data you give it, it synthesizes beautifully using the keywords you provide*." (Project 2, Facilitator)

Despite a better-structured process for using GenAI, managing the large volumes of data and diverse prompts in parallel with coordinating the workshop activities remained challenging. Continuously updating and maintaining data quality throughout the sprint also posed difficulties.

These challenges highlighted the need for a new method to integrate contextual inputs more seamlessly and systematically during the design sprints.

In November 2023, just a few weeks before the subsequent sprint, the launch of the GPT store presented an opportunity to refine the use of GenAI. This new resource allowed for the creation of customized Large Language Models[2], providing a way to transition from using a basic prompt bank in a document to employing reusable GPTs for generic tasks within the sprint. The facilitator would be able to use custom-made GPTs both for direct interactions during workshops and for background tasks between sessions. This change could potentially reduce his cognitive efforts and minimize errors associated with manually managing prompt templates and data.

**Project #3: Building and Leveraging Reusable Capabilities through GPTs**

The third innovation project took place at a public television company focused on educational and cultural content, intending to develop a new digital experience for young consumers. This sprint demonstrated the most significant evolution among the three projects, with the introduction of custom-made GPTs to address the previously experienced inefficiencies associated with working with a prompt repertoire – a "toolbox, but a messy one" in the facilitator's words (a notable lesson from Project #2). To optimize his use of GPTs, the facilitator developed a dynamic *project context document*—a structured compilation of information generated by both participants and GenAI throughout the sprint. This document acted as a collective memory, cataloging key information like business objectives, user requirements, success criteria, opportunity areas, and solutions. It was systematically updated with new data inputs from workshop exercises to ensure GenAI outputs remained relevant and well-aligned with the sprint's progress. In parallel, the facilitator worked on the transparency of GenAI-generated content by clearly tagging each piece of information presented to the team with its origin—human or machine—and categorized as 1. Untouched, 2. Reformulated by AI, or 3. Generated by AI. This approach aimed to maintain trust and acceptance by clearly indicating the source of inputs and level of AI involvement in the sprint. It also facilitated the evaluation of how AI-generated ideas were received and valued by the team.

**Project framing** and **understanding** continued to be carried out through research, surveys, and individual meetings. This was followed by the *alignment workshop* (**defining**), in which a fine-tuned and streamlined approach was adopted. The prompts repertoire used in the previous project was transformed into specialized GPTs. Each GPT was tailored to a specific task within the sprint. For example, one of them, named "Them-inator", was tasked with extracting themes from a large set of human inputs. The output could then be used in "affinity clustering" exercises, where team members work on how diverse ideas are associated, developing a deeper understanding of the problems. The workshop ended with the team selecting the specific problem to pursue.

[2] https://openai.com/index/introducing-gpts/

The **imagining** stage mirrored that of Project #2 but included an enhanced use of the text-to-image GenAI tool, DALL-E within the workshop. This tool generated visual representations of ideas suggested by the team, stimulating inspiration and creativity.

**Prioritization** occurred over three intensive *three-hour workshops*. Initially, each team member picked their top ideas (based on their ability to meet the predefined success criteria) and designed their solution concept. Additionally, the facilitator created an AI-generated concept using one of his specialized GPTs. In the following workshop, team members discussed the human-generated concepts and voted on the most promising one. Noting that the AI-generated concept closely resembled one proposed by a team member, the facilitator opted not to present it to avoid potential voting bias from duplicate propositions. This stage of the sprint typically exhausts participants due to demanding deliberation. To alleviate fatigue (and maximize productivity), the facilitator used a GPT to convert images of the selected concept into text, incorporating it into the project context document. This document then helped another GPT to formulate hypotheses for user testing. In the final workshop, where a considerable amount of information needs to be converted into more tangible artifacts for user testing, the facilitator pre-generated two baseline user test flows using another set of specialized GPTs. That way, the team was able to concentrate its energy on refining rather than creating test flows from scratch, streamlining the final decision-making process.

During the **validation** phase, the facilitator used DALL-E to create a cartoon-style prototype (the target users were children, making this format well-suited). Team members individually reviewed this prototype and provided feedback within a day. The final prototype, again generated with the help of a GPT that structured the inputs for DALL-E, was then user-tested. User feedback was recorded, automatically transcribed using a speech-to-text tool, and cleaned with a GPT. Finally, another GPT helped synthesize, standardize, and organize the data for communication to the team.

**Learnings from Project #3** – The integration of GenAI in this third project showcased its transformative impact on storyboarding and prototyping. Participants were better able to visualize potential solutions, which enhanced idea generation and accelerated the convergence phases.

> *"The facilitator used GenAI to generate images to illustrate the concepts. We have UX designers or a digital art director; they could have done it. But quickly testing an idea allowed us to save time and efficiency. (...) It also allowed us to quickly test [the prototype] with our target population, in this case, young people, and get quality feedback" (Project #3, Team member #1).*

Importantly, transparent and controlled use of GenAI boosted team members' confidence in the process.

> *"I really appreciated the transparency, both for the text and the images. If that had not been the case and we had selected an AI-generated idea during the decision phase, I think my trust would have been affected. (...) The visuals, if I hadn't been told they were AI-generated, would have bothered me less, but if it was about central ideas, I would have found it unfortunate." (Project #3, Team member #1)*

> *"For exploratory phases, I don't see a problem with AI. It would even be well-regarded. However, for pivotal phases like making decisions or voting on something important like a success factor, there would be rejection if we relied solely on AI to determine what is important. We are better positioned to know. We are the employees who will execute the final solution." (Project #3, Team member #1)*

Furthermore, the use of the project context document reinforced its role as a key reference point and collective memory, where all information and ideas are dynamically gathered, refined, and organized. This document served as a foundation for team alignment and shared understanding, helping to prevent deviations from agreed-upon problems and decisions.

> *"Sometimes we had a lot of ideas, and the AI summarized them into a concise text. This allowed us to wrap our heads around it. It gave us a base; there were many fragmented ideas that everyone had contributed, and now we had a text on which we all agreed. This way, we started from the same base to move forward. We could always refer to that" (Project #3, Team member #1).*

Figure 3 depicts a high-level view of the facilitator's evolving use of GenAI across three sprints. Project #1 began as a rudimentary exploration of ChatGPT's capabilities, focusing on a few tasks with contextual information drawn mostly from the first two phases of the DS. In Project #2, contextual data was consolidated into a more structured document, enhancing the organization and use of data and prompts during workshops. GenAI's role shifted from being manipulated "like a puppet" by a co-facilitator to being directly utilized by the facilitator during and between sessions. Project #3 marked further advancements, with the deployment of multiple GPTs tailored to specific “jobs to be done” within the sprint, supporting information management and processing. A dynamic project context document ensured alignment between workshop activities and the generated data. The two subsequent figures offer a more granular view of the key design practices deployed across projects as well as across phases. Figure 4 highlights key practices within the problem space while Figure 5 focuses on the solution space.

**PROJECT #1**
**GEN AI as a PUPPET**

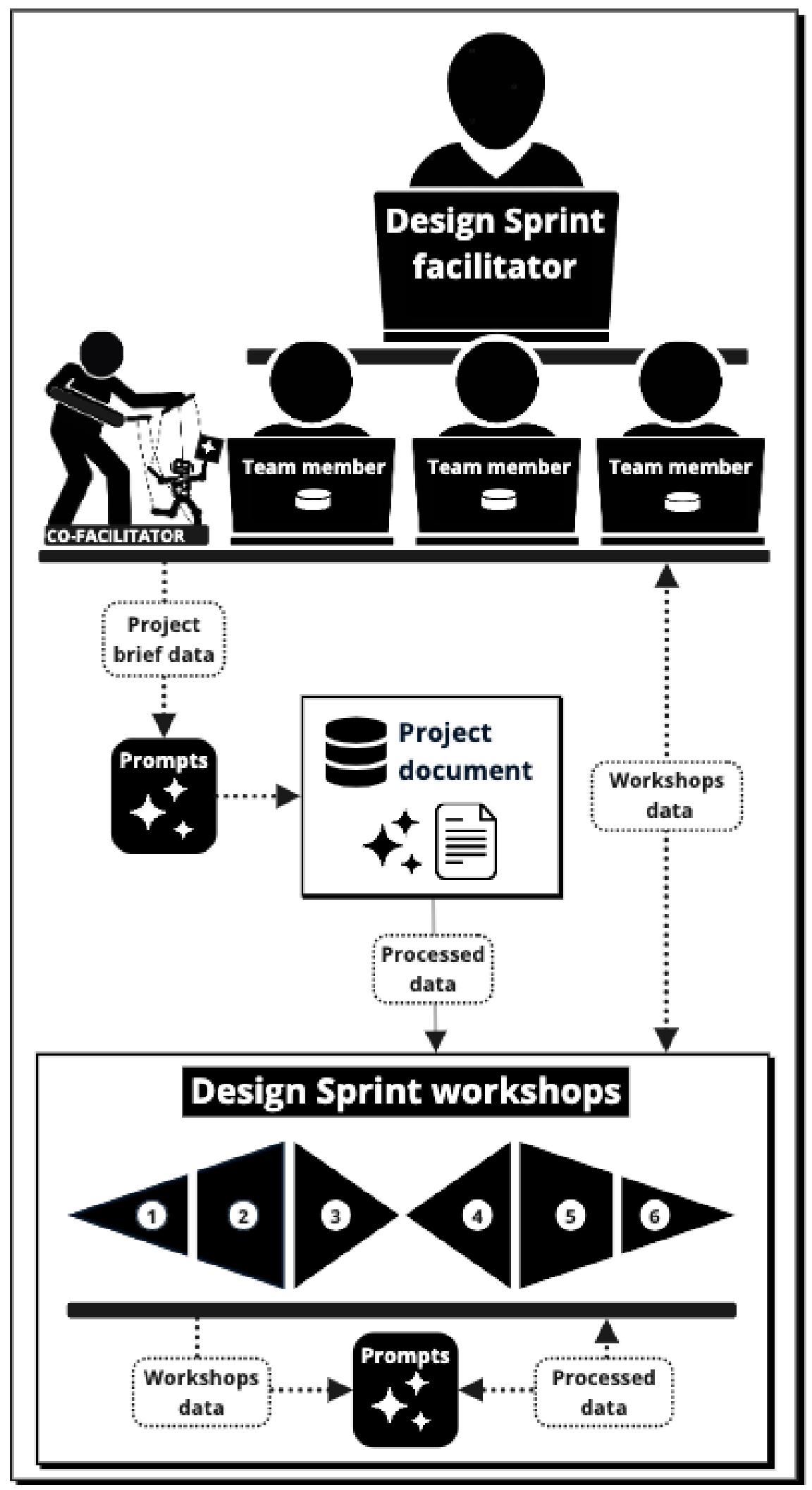

**PROJECT #2**
**GEN AI as a TOOL BOX**

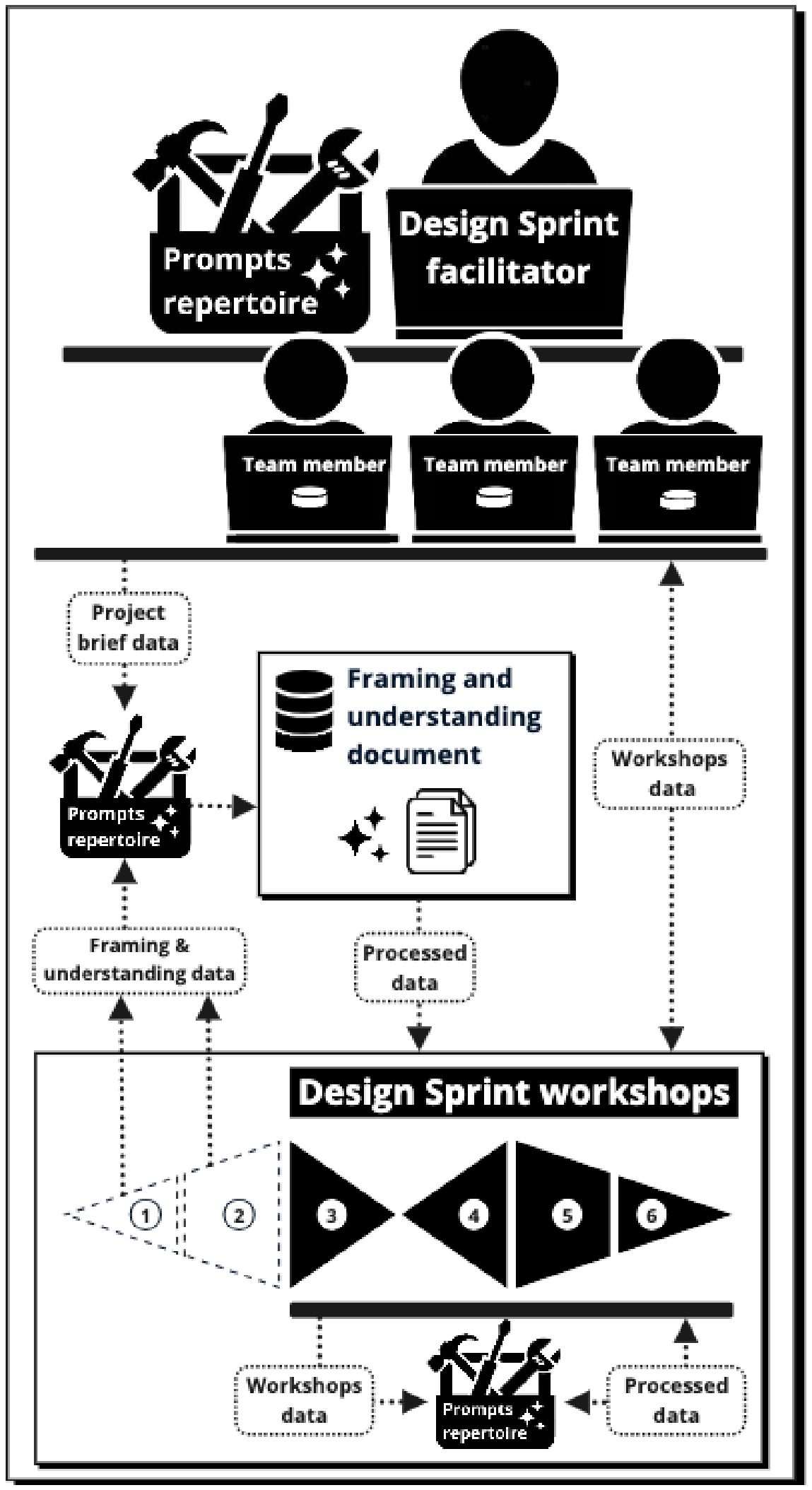

**PROJECT #3**
**GEN AI as a GPTs LIBRARY**

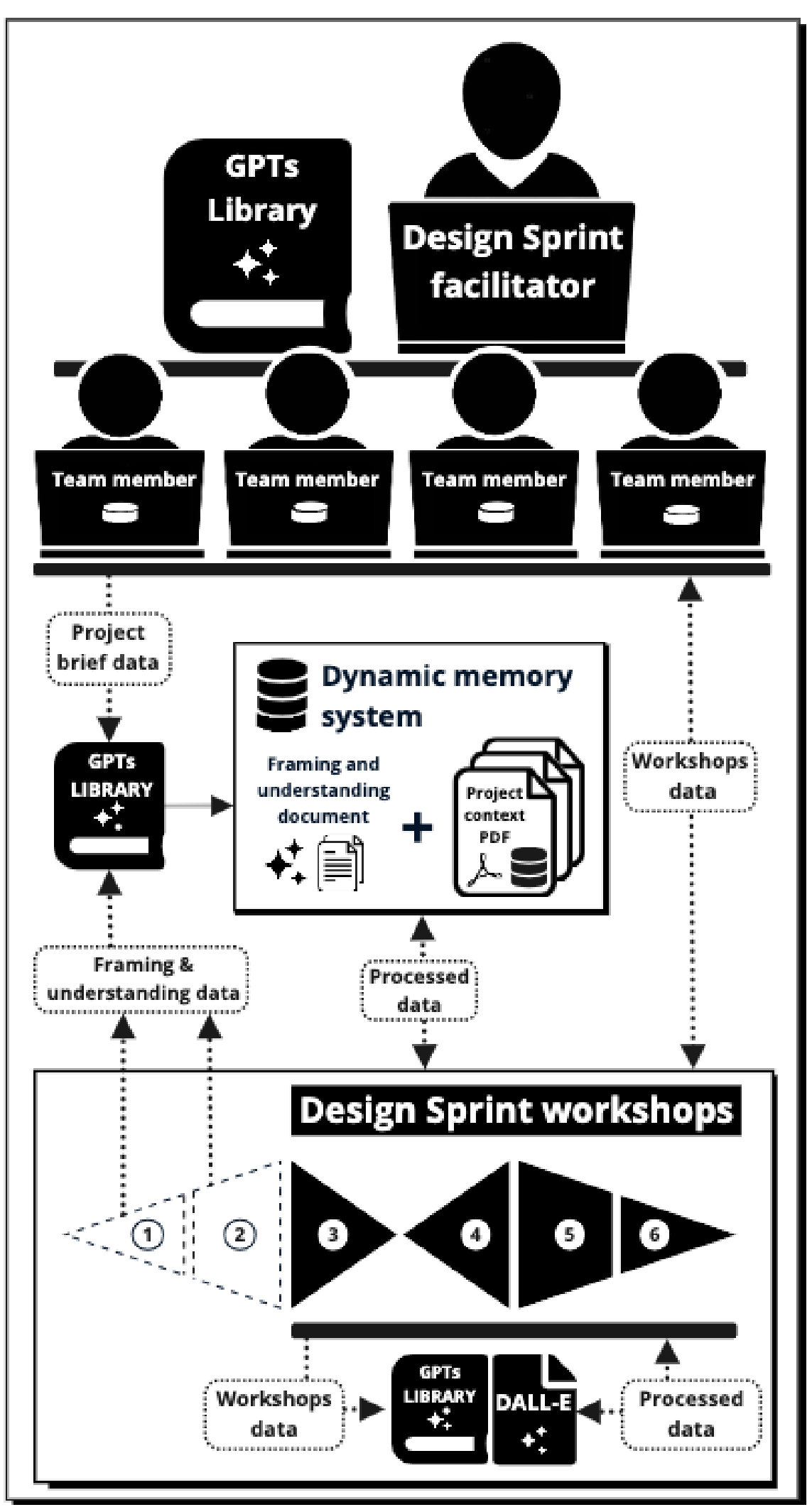

Figure 3. Summary Architecture of GenAI's Use Across Projects

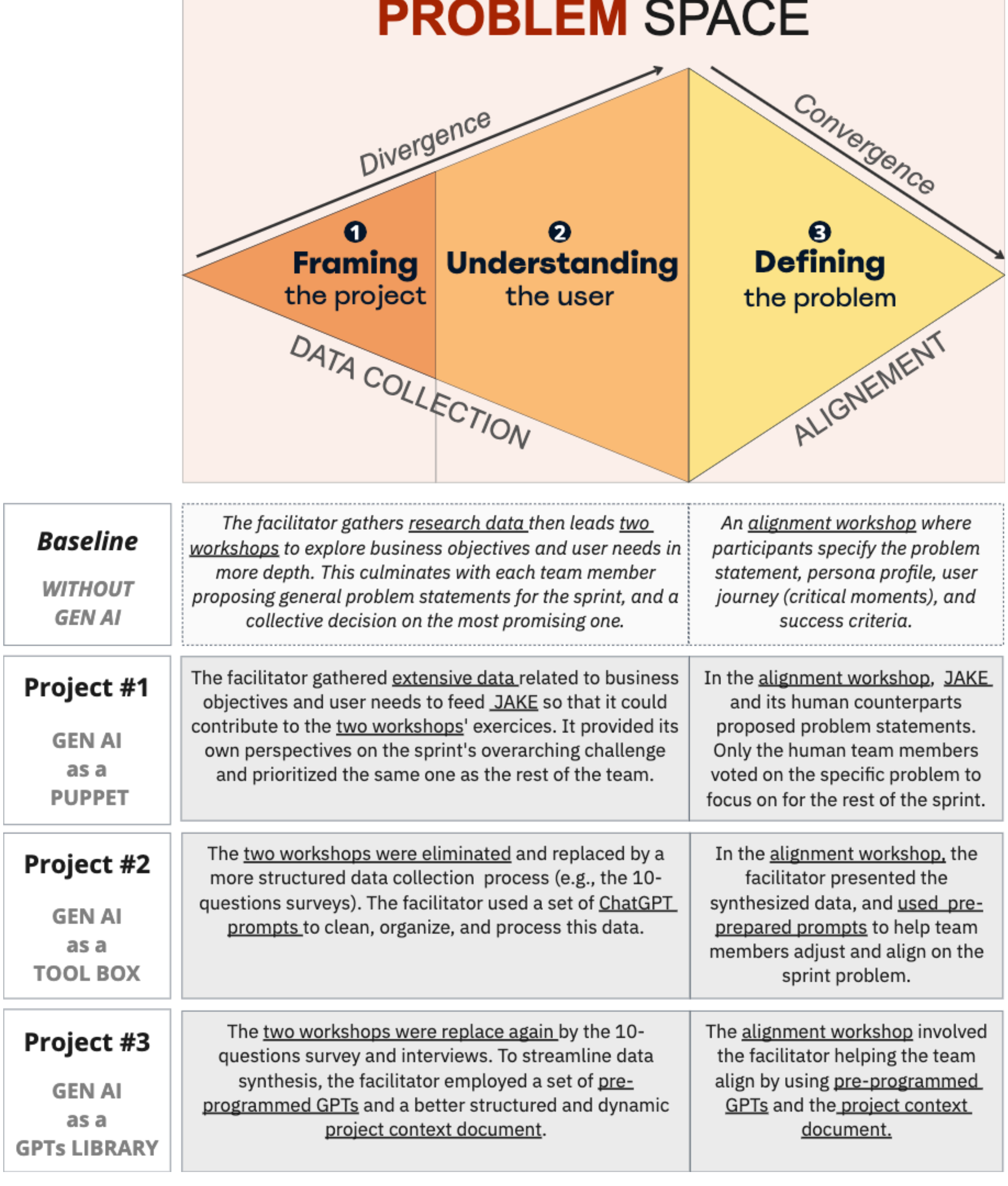

| | Framing the project / Understanding the user | Defining the problem |
|---|---|---|
| ***Baseline*** *WITHOUT GEN AI* | *The facilitator gathers research data then leads two workshops to explore business objectives and user needs in more depth. This culminates with each team member proposing general problem statements for the sprint, and a collective decision on the most promising one.* | *An alignment workshop where participants specify the problem statement, persona profile, user journey (critical moments), and success criteria.* |
| **Project #1** GEN AI as a PUPPET | The facilitator gathered extensive data related to business objectives and user needs to feed JAKE so that it could contribute to the two workshops' exercices. It provided its own perspectives on the sprint's overarching challenge and prioritized the same one as the rest of the team. | In the alignment workshop, JAKE and its human counterparts proposed problem statements. Only the human team members voted on the specific problem to focus on for the rest of the sprint. |
| **Project #2** GEN AI as a TOOL BOX | The two workshops were eliminated and replaced by a more structured data collection process (e.g., the 10-questions surveys). The facilitator used a set of ChatGPT prompts to clean, organize, and process this data. | In the alignment workshop, the facilitator presented the synthesized data, and used pre-prepared prompts to help team members adjust and align on the sprint problem. |
| **Project #3** GEN AI as a GPTs LIBRARY | The two workshops were replace again by the 10-questions survey and interviews. To streamline data synthesis, the facilitator employed a set of pre-programmed GPTs and a better structured and dynamic project context document. | The alignment workshop involved the facilitator helping the team align by using pre-programmed GPTs and the project context document. |

Figure 4. Key Characteristics of the Design Practice Across Projects and Phases (problem space)

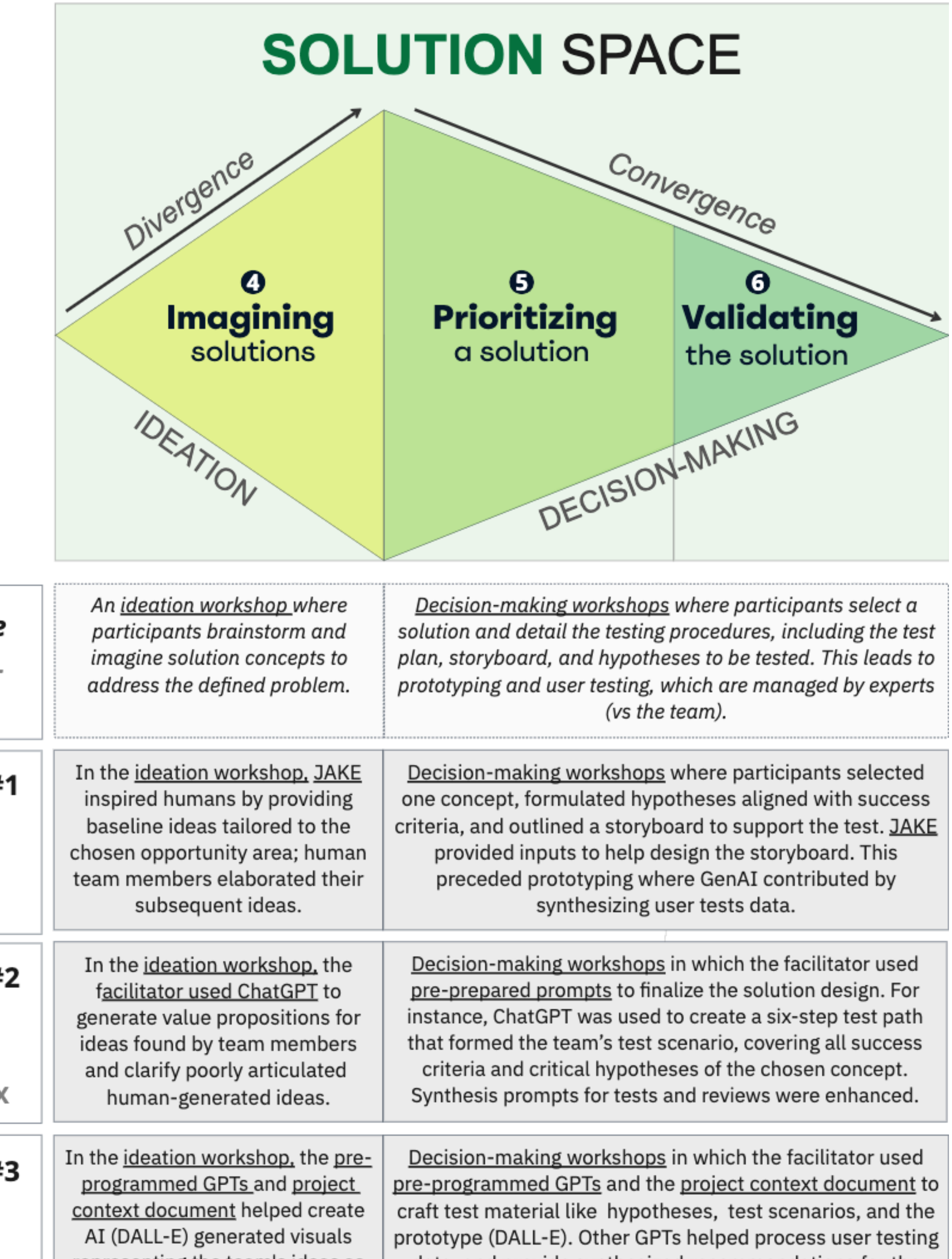

| ***Baseline*** *WITHOUT GEN AI* | *An ideation workshop where participants brainstorm and imagine solution concepts to address the defined problem.* | *Decision-making workshops where participants select a solution and detail the testing procedures, including the test plan, storyboard, and hypotheses to be tested. This leads to prototyping and user testing, which are managed by experts (vs the team).* |
|---|---|---|
| **Project #1** GEN AI as a PUPPET | In the ideation workshop, JAKE inspired humans by providing baseline ideas tailored to the chosen opportunity area; human team members elaborated their subsequent ideas. | Decision-making workshops where participants selected one concept, formulated hypotheses aligned with success criteria, and outlined a storyboard to support the test. JAKE provided inputs to help design the storyboard. This preceded prototyping where GenAI contributed by synthesizing user tests data. |
| **Project #2** GEN AI as a TOOL BOX | In the ideation workshop, the facilitator used ChatGPT to generate value propositions for ideas found by team members and clarify poorly articulated human-generated ideas. | Decision-making workshops in which the facilitator used pre-prepared prompts to finalize the solution design. For instance, ChatGPT was used to create a six-step test path that formed the team's test scenario, covering all success criteria and critical hypotheses of the chosen concept. Synthesis prompts for tests and reviews were enhanced. |
| **Project #3** GEN AI as a GPTs LIBRARY | In the ideation workshop, the pre-programmed GPTs and project context document helped create AI (DALL-E) generated visuals representing the team's ideas as well as one of its own. | Decision-making workshops in which the facilitator used pre-programmed GPTs and the project context document to craft test material like hypotheses, test scenarios, and the prototype (DALL-E). Other GPTs helped process user testing data, and provide synthesized recommendations for the review. |

Figure 5. Key Characteristics of the Design Practice Across Projects and Phases (solution space)

# 4. Recommendations

The insights gained from our research led us to outline four key recommendations for effectively integrating GenAI into collaborative, human-centered innovation processes.

## 4.1 Establish a Collaborative Intelligence Value Loop with GenAI

**Enable value co-creation in task distribution**. Integrating GenAI in a human-centered innovation process like design sprints demands an understanding of how to co-create value, harnessing both human and AI capabilities. In this regard, our research underscores human team members' indispensable contributions: providing contextual information, validating GenAI outputs, and making final decisions. Humans offer unique contextual insights, capturing nuances that GenAI is likely to miss. In the framing stage of the DS we studied, team members provided insights about what they thought were the firm's challenges and the client's needs. Without this input, GenAI might generate innovative but irrelevant or inappropriate ideas for the target audience. Additionally, human validation is crucial to ensure the accuracy and applicability of GenAI-generated content, filtering out impractical or less effective solutions. In our projects, GenAI could produce scenarios and prototype drafts to validate the concept, but the team had to elevate and refine these to meet project objectives. In brief, human decision-making is vital for choosing the best ideas and strategies, considering feasibility, ethical implications, and long-term impact. During the convergence phase of the sprints, only human team members voted on the most promising opportunities, ensuring that decisions aligned with the company's culture and strategic goals.

Consistent with emerging research (Alavi & Westerman, 2023; Coombs et al., 2020; Richardson, 2020), our findings also suggest that GenAI excels at foundational tasks like reformulating, summarizing, and categorizing, all critical in collaborative innovation. By assigning these tasks to GenAI, participants can focus more on activities for which nuance and context are key (Jia et al., 2024), like discussing and evaluating ideas, leading to more informed and confident decisions. While GenAI can certainly help the process of developing a collective understanding of a problem and push teams to explore a broad array of initial possible solutions, its greatest value emerges in convergence when it is fuelled by humans' contextual insights emerging during divergence. Table 1 presents a synthesis of how tasks can be effectively distributed between team members and GenAI throughout design sprint phases, highlighting GenAI's value addition.

Table 1. Collaborative Roles of GenAI and Humans Across Design Sprint Phases

| **Sprint Phases** | **Humans' Role** | **GenAI's Role** | **Added value of GenAI** |
|---|---|---|---|
| Framing the Project | - Produce contextual insights via surveys and interviews.<br>- Validate business goals. | - Summarize and structure the insights from surveys and meetings into a summary document. | - Optimize a labor-intensive information synthesis process for the facilitator. |
| Understanding the User | - Show empathy through various techniques and tools to gather user needs. | - Consolidate team member perspectives from surveys and meetings into a single context document | - Facilitate a clearer, shared understanding of opportunities.<br>- Facilitate the sharing of user frustrations, needs, and aspirations within the team |
| Defining the Problem | - Consolidate understanding data (pre-workshop).<br>- Identify complex issues within the project.<br>- Select the most promising opportunity areas. | - Generate balanced "How might we…" exploration questions<br>- Organize team member contributions into thematic groups.<br>- Propose a set of problem statements that meet the sprint's success criteria. | - Lighten the categorization burden for team members.<br>- Ensure equal consideration of team member's contributions.<br>- Facilitate team alignment during workshop exercises. |
| Imagining Solutions | - Envision innovative solutions to achieve business goals while meeting user needs. | - Inspire the team with examples tailored to the selected opportunity area.<br>- Refine and simplify the team's proposed solution statements. | - Propel the team toward more innovative solutions.<br>-Help visualize potential solutions.<br>- Enhance the clarity and impact of envisioned solutions. |
| Prioritizing a Solution | - Refine, organize, and assess the impact of imagined solutions.<br>- Vote on the best solution concept for testing.<br>- Identify and articulate critical hypotheses to test based on the most popular concept. | - Extract solution descriptions generated by the team.<br>- Uncover patterns and connections between ideas.<br>- Enhance and validate statements of critical hypotheses for testing.<br>- Develop detailed test scripts for the user experience to be tested. | - Improve the clarity of solution descriptions.<br>- Standardize concept presentations.<br>- Salvage good but poorly articulated ideas.<br>- Facilitate collective sensemaking.<br>- Support storyboard development.<br>- Align the team on hypotheses to test |
| Validating the Solution | - Review and select innovative solutions for the prototype.<br>- Provide and receive feedback on the prototype's value.<br>- Conduct and assist with user testing.<br>- Learn from and act based on results. | - Create prototype wireframes.<br>- Generate value proposition texts for the presented solution.<br>- Transcribe and synthesize user interviews.<br>- Extract key observations. | - Accelerate convergence.<br>- Ensure consistency in prototype design.<br>- Streamline the development of test materials and data analysis.<br>- Clarify and standardize sprint review recommendations. |

**Define and communicate roles clearly**. However, we found that roles must be clearly defined and communicated to the team. A clear distinction between human and AI contributions helps avoid confusion and highlights the distinctive expertise of the participants. In two instances, the distinction between roles was blurred, which triggered reactions from the facilitator or the participants. In the first project, GenAI was anthropomorphized and treated as a participant, although it did not have any decision-making control. Even though GenAI was framed as a "contributor" rather than a decision-maker to clarify roles and responsibilities, the facilitator later replaced it with a human member to enrich the group with more contextual knowledge about the problem and context. The facilitator valued human contributions over AI and felt it was necessary to ensure the integrity of the innovation process. Neglecting these role distinctions can have significant consequences. In the third project, some participants expressed resistance to outcomes that were generated solely by AI during pivotal phases. This reaction showed the importance of human control in critical stages, as people are ultimately responsible for the final solution and must champion its relevance to their colleagues after the sprint.

**Continuously assess GenAI's role and impact**. The ongoing assessment of GenAI's role and its impact on both team efficiency and the human psyche was essential in supporting the innovation process. For instance, in the third project, the facilitator chose not to unveil the GenAI-generated concept during the prioritization phase, as it was similar to another concept proposed by a team member, following a precautionary principle. Across the projects, the facilitator's reflexive approach, along with the documentation of successes and failures, allowed him to adapt his methods to protect participants' engagement, sense of ownership, and creativity while leveraging GenAI. Practitioners should adopt a similar approach of continuous improvement and exercise caution when bringing GenAI into the innovation process.

### 4.2 Build Trust in GenAI

Incorporating GenAI into an innovation process requires helping the team build trust in the AI-augmented process, ensuring they feel secure and confident in the technology's capabilities and use. Transparency is critical in fostering this trust (Glikson & Woolley, 2020; Wanner et al., 2022). Like many algorithmic technologies, GenAI tools often operate as black boxes (Pasquale, 2015), obscuring the origins and processes behind both the input they receive and the information they generate. To address this opacity, practitioners must proactively address how these tools are deployed within their teams.

**Clearly explain GenAI usage**. A first step in building trust is to explain clearly when, how, and for what purposes GenAI is being used. Practitioners can provide explicit details about instances of GenAI-generated content, the tasks it supports, and the outcomes it aims to achieve. For example, in our study, the facilitator met each team member before the start of the DS to explain how GenAI would be used and address any possible concerns.

**Demonstrate GenAI through use cases.** Another leverage to build trust is to use concrete examples to help the team better appreciate not only the advantages of GenAI but also its limitations. These conversations can take place during individual meetings before gathering the team into workshops in order to identify and address reluctance before participants engage in the activity. In our studied projects, this approach helped demystify the technology and promote informed reliance on GenAI (Teodorescu et al., 2021), reducing potential skepticism and fostering a culture of openness.

**Label all GenAI-generated content**. Another effective action in building trust through transparency is explicitly labeling AI-generated ideas and content. GenAI-generated content is judged more acceptable during the creative process when clearly acknowledged (Doshi & Hauser, 2023). When team members know certain pieces of content or suggestions are produced by GenAI, they can critically evaluate the information. This practice not only maintains clarity but also prevents mistakenly attributing human characteristics or intentions to AI outputs. For example, in the third project, the facilitator clearly annotated, throughout the entire design process, whether the data was created by humans, reformulated by GenAI, or fully generated by GenAI. This clear understanding of the source of the content avoided confusion. It allowed team members to modify or refine the AI-generated material as necessary, enhancing their engagement with it. Crucially, this also empowered the team to feel more in control of the innovation process

### 4.3 Develop Robust Data Collection and Curation Workflows

Practitioners integrating GenAI into collaborative human-centered innovation must invest significant effort in data management. Indeed, the old adage “garbage in, garbage out” resonates strongly with the use of GenAI in design sprints (Kilkenny & Robinson, 2018). High-quality data inputs are essential to generating better outputs that enhance the problem-solving process. This became evident in the first project, prompting the facilitator to prioritize the development of a more effective data architecture (see Figure 2).

**Collect detailed, contextually relevant data**. Effective collection, cleaning, and structuring of contextual and human-generated information are crucial for leveraging GenAI's capabilities. Specialized prompts designed for processing this data are also essential to maximize GenAI's potential. In preparation for a design sprint, the facilitator in our projects collected extensive information about the firm's challenges through surveys and interviews. GenAI was invaluable in synthesizing the diverse perspectives among team members. It is important to note that managing this data also includes manually reviewing and occasionally correcting GenAI outputs to ensure their accuracy and clarity.

**Develop a "living document" of contextual data**. Maintaining constant alignment between GenAI and the team is crucial for achieving synergistic outcomes from their interactions (Han et al., 2024). In our study, alignment was facilitated by a "project context document" (project #3) that compiled all relevant data. This document was regularly updated and synchronized with inputs from pre-DS surveys and information generated during each DS activity. Such alignment ensures that GenAI's outputs remain relevant and accurate. The project context document can serve as a living memory of the innovation process and should be integrated into each prompt to provide context for the GenAI tool. Structuring the document with sections, sub-sections, and tagging each piece of information according to its type reduces GenAI's hallucination rate and facilitates easy retrieval. While creating, updating, and curating this document demands substantial effort, it significantly improves the quality and relevance of GenAI's tasks during DS activities, and specialized GPTs can assist with this task.

**Implement systematic data workflows**. The structuring of a data management architecture can take different forms. In our study, significant progress in that regard was achieved in Project #3 with the development and use of a library of ready-made GPT capabilities together with the output templates and workflows associated with the various activities of the DS. When determining tasks for GenAI, design sprint facilitators need to consider three essential components: 1) the prompt itself, detailing the task; 2) the contextual data needed from the project context document; and 3) reference data from relevant documents, such as previous examples or output formats.

### 4.4 Embrace a Craftsman's Discipline

GenAI-augmented innovation is a craft demanding "care and ingenuity and requires patience and perseverance" (Sennett, 2008). A final recommendation for practitioners integrating GenAI into their innovation processes is thus to adopt a craftsman-like approach, which involves a commitment to continuous learning (honing skills regularly), curiosity (remaining updated on the latest technological developments), ingenuity (approaching challenges with a problem-solving mindset), and contextual sensitivity (demonstrating adaptability).

**Promote continuous exploration and improvement.** GenAI's application in innovation processes is still being explored, and no definitive template exists to standardize these methods fully. As GenAI technology progresses, practitioners must be prepared to encounter and learn from mistakes or inefficiencies, continuously refining their methods. This includes enhancing prompts and improving data management techniques during and between design sprint workshops. For example, the facilitator in our study regularly updated prompting strategies, explored new use cases, and engaged with advanced techniques like chain-of-thought prompting (Wei et al., 2023) to enhance the usability and utility of GenAI outputs. While this ongoing experimentation may be laborious at times, it is essential to maximize the effectiveness of GenAI in the innovation process. GenAI is a flexible technology. As such, its exploration within the context of design sprints, through trial and error, can lead to significant improvements (Schmitz et al., 2016). Therefore, keeping pace with the latest advancements in GenAI technology and discovering its potential applications is crucial for practitioners to fully exploit GenAI's capabilities in supporting innovation.

**Tailor GenAI integration to fit the context**. Practitioners must recognize the value GenAI can bring to innovation processes but also be discerning about its appropriateness in specific contexts. Each organizational and team setting varies, requiring personalized approaches to the integration of technology. The team shapes and may ultimately adopt the final solution, necessitating a flexible mindset to meet its specific needs. For instance, while GenAI's assessments might align well with the team's important choices during the sprint (e.g., selecting the problem to solve or the solution to design), as we observed in our projects, limiting its use might still be wise to ensure team members maintain ownership of the chosen problem or solution. In other words, despite GenAI's impressive automation capabilities, they may not always be suitable or desired in every scenario.

Table 2 summarizes our four key recommendations together with enabling actions and their corresponding rationale.

Table 2. Recommendations for Effectively Integrating GenAI into Collaborative Human-Centered Innovation Processes

| Recommendations | Enabling actions | Rationale |
|---|---|---|
| Role Distribution: Establish a Collaborative Intelligence Value Loop with GenAI | Enable value co-creation in task distribution. | GenAI and humans have complementary capabilities. By assigning demanding information processing tasks to GenAI, humans can focus more on nuanced, context-dependent activities. |
| | Define and communicate roles clearly. | Clear role definitions avoid misunderstandings as well as emphasize the value of human contributions and the complementarity role of GenAI. |
| | Continuously assess GenAI's role and impact. | Regular evaluation ensures GenAI enhances the innovation process without undermining human engagement, ownership, and creativity. |
| Team Management: Build Trust in GenAI | Clearly explain GenAI usage. | Transparency builds trust. Clear explanations of GenAI's usage demystify stereotypes (utopian or dystopian), reduce skepticism, and improve buy-in. |
| | Demonstrate GenAI through use cases. | Concrete examples help team members understand GenAI's potential and limitations, reducing anxiety and encouraging willingness to explore the new practice. |
| | Label all GenAI-generated content. | Labeling helps team members critically evaluate outputs and understand GenAI's role, promoting trust and informed decisions. |
| Data Management: Develop Robust Data Collection and Curation Workflows | Collect detailed, contextually relevant data. | High-quality data gives GenAI the context it needs to generate valuable outputs. The richer the data, the better the results. |
| | Develop a "living document" of contextual data. | A dynamic document ensures GenAI has the latest insights and helps keep the project aligned with evolving understanding and objectives. |
| | Implement systematic data workflows. | Pre-defined workflows and templates ensure consistent, efficient use of GenAI in the design sprint. |
| Mindset: Embrace a Craftsman's Discipline | Promote continuous experimentation and improvement. | GenAI and best practices for using it evolve quickly. Adapting strategies based on evolving capabilities allows organizations to take advantage of new features and identify better ways of doing things. |
| | Tailor GenAI integration to fit the context. | Teams need to "own" the problem they work with as well as the solution they test. Thus, their sensitivity to GenAI use and inputs deserves primary consideration. |

## 5. Final Reflections: Practical Applicability and Future Developments

We expect our four recommendations to be applicable across organizational environments—from startups to multinational corporations. They should also be compatible with different innovation approaches, including Agile, Lean, and various adaptations of applied Design Thinking methods. Certain factors may, however, either facilitate or hinder the effective deployment of these recommendations. Moreover, when it comes to GenAI's capabilities for innovation, we are still in the early stages. Thus, we now discuss applicability constraints as well as what comes next at the intersections of innovation, creativity, and GenAI.

## 5.1 How Does this Apply to Your Context?

It is well-established that an *organization's culture* significantly influences its ability to innovate. Organizations that cultivate a culture of experimentation (Thomke, 2020) and agility (Rigby et al., 2016) are typically better positioned to adopt new approaches and technologies within their operations (Naranjo-Valencia et al., 2011). In contrast, organizations with more rigid cultures may face greater challenges, requiring more time and effort to overcome resistance to change. Interestingly, our research contrasts with these traditional views. We examined projects at three organizations, varying in their appetite for innovation. The first was a small private entity deeply immersed in a culture of innovation and creativity, making it an ideal setting for exploring GenAI's capabilities in design sprints. The other two projects took place in environments much less inherently tied to an innovation culture: one within a major metropolitan municipality and the other at a national public media organization. Despite this difference, we did not observe any significant barriers as a result of organizational culture. All team members actively and positively engaged with the new approach, suggesting that GenAI can be effectively applied within innovation processes across a broad range of organizational contexts.

One factor that we expect to more fundamentally influence the successful implementation of our recommendations is *an organization's ability to source an effective facilitator*. Prior research, such as that by Gotlar and Hutley (2024), suggests that this individual should ideally be external to the team or organization. Our findings also underscore the importance of this role, but further highlight key traits such as empathy, innovativeness, meta-knowledge, and ambidexterity. *Empathy* is about understanding (Ventura, 2018, p. 201). It is both cognitive (i.e., the ability to understand others' perspective) and emotional (sharing feelings) (Montonen et al., 2014). Our projects' facilitator effectively communicated the benefits and risks of GenAI, attentively addressing team members' concerns. This was crucial for nurturing trust and psychological safety within the teams, both essential components in the high-pressure environment of a design sprint (Clark, 2022).

*Innovativeness* relates to using digital tools in novel ways to enhance task performance. The importance of individual innovativeness in facilitating IT-based innovations is well-documented (Abubakre et al., 2020; Jackson et al., 2013; Yi et al., 2006) and was apparent in our study as the facilitator continually challenged their use of GenAI, restructuring workshop activities and testing new practices, which significantly advanced data management architecture and task distribution. "*Meta knowledge*" involves understanding the synergies between GenAI and human team members (Fugener et al. 2022). This knowledge is essential for identifying opportunities and avoiding pitfalls within the "collaborative value loop." The facilitator in our projects developed meta-knowledge by reflecting on which tasks ("jobs to be done") were best suited for himself, the AI, the team, or a combination thereof. *Ambidexterity*, defined as the ability to exploit existing competencies and explore new opportunities simultaneously (March, 1991; Raisch et al., 2009), is another vital trait for facilitators working with GenAI in innovation projects. Facilitators face challenging decisions in allocating resources (time and attention) between exploitation, which refines existing processes, and exploration, which seeks new possibilities. It is essential to maintain a balance, fostering incremental improvements while also investigating more transformative opportunities (O'Reilly & Tushman, 2004; Tushman & O'Reilly, 2002). In our study, the facilitator adeptly managed this balance. He saw each new project as an occasion to refine existing ways of using GenAI while staying alert to opportunities that entirely new architectures could offer (e.g., GPTs). This dual focus allowed him to enhance his data management efforts, leveraging both established (exploitation) and innovative (exploration) approaches.

Finally, we expect our practical recommendations to be largely adaptable to various *innovation formats*. Design sprints offer a flexible framework that can be operationalized with varying focus on different phases of the process. Our research shows that GenAI can significantly enhance all phases, from problem framing and ideation to prioritization and validation. However, it also highlights that GenAI's effectiveness is contingent upon possessing substantial contextual knowledge about the problem at hand, in other words, training the AI (Gotlar and Hutley 2024). Therefore, shorter sprint formats, which may skimp on extensive project framing and the gathering of contextual data, could see diminished benefits from GenAI integration. Moreover, while our study primarily focuses on design sprints, the insights and recommendations are also applicable to other innovation approaches, such as Agile or Lean., which prioritize rapid iteration, customer feedback, and a data-driven approach to decision-making (Rigby et al., 2016). For instance, Project

#3 illustrated how GenAI could be a particularly effective catalyst for rapid prototyping (Parra Pennefather, 2023).

## 5.2 Practical Challenges and Considerations

While using GenAI in innovation holds promise, it remains in its early stages, presenting practical challenges and considerations. Moreover, the long-term implications of both the technology and its use are not yet fully understood, and further exploration is needed to anticipate both the potential risks and opportunities. Thus, integrating GenAI into the innovation process will demand thoughtful, ongoing engagement from practitioners.

**Avoiding Overreliance on AI.** Mindlessly accepting GenAI suggestions can lead to suboptimal outcomes in the long term. The success of team innovations relies on the group's collective intelligence, derived from diversity and the unique competencies of its members (Polzer et al., 2002). In design sprints, teams are intentionally diverse, encompassing various professional backgrounds, hierarchical levels, and departmental roles to effectively engage with the four critical innovation factors—desirability, feasibility, viability, and adaptability (Osterwalder & Pigneur, 2010). Each team member provides distinct and complementary insights crucial for problem-solving. Integrating GenAI is like adding specialized collaborators (Anthony et al., 2023), who assist in tasks such as synthesizing or clarifying, and can also take more active roles like inspiring or prioritizing. This necessitates careful orchestration to ensure all contributors, human or AI, effectively utilize their unique capabilities.

Consequently, facilitators must develop mechanisms to ensure balance, as recent evidence suggests that overreliance on AI can undermine individuals' sense of skills (Strich et al., 2021), stifle creativity and innovation (Rinta-Kahila et al., 2023) or inhibit the intelligence of the group by homogenizing its members' contributions (Burton et al., 2024). Nonetheless, AI-centric approaches like the "Stingray model" by Board Innovation[3] are emerging from practice, posing interesting dilemmas regarding the style of collaborative intelligence to adopt with GenAI and its potential consequences on the innovation process. Furthermore, similar to mimicking behaviors in a simulacrum of human society (Park et al., 2023), there is an ongoing discussion about whether synthetic users should be employed in user experience research (Li, 2024; Rosala & Moran, 2024).

[3] https://www.boardofinnovation.com/blog/the-ai-powered-stingray-model-innovation/

It raises additional concerns about the extent to which GenAI should be integrated, particularly in the data collection process. In such cases, practitioners must remain mindful of the potential unforeseen consequences of these integrations.

**Leveraging GenAI's Multimodal Capabilities.** The diversity of interaction modes offered by GenAI can enhance team creativity (Zhang et al., 2024). Our innovation projects primarily utilized text-based tools, consistent with prevailing GenAI applications. Yet, the shift towards multimodal GenAI—integrating multiple input types (or modes) like text and images to generate varied outputs including text, images, videos, and audio—is emerging as a significant trend (Boston Consulting Group, 2023). Experts from a recent Delphi study anticipate that multimodal GenAI will more positively impact innovation performance and competitive advantage in adopting firms compared to unimodal systems (Mariani & Dwivedi, 2024). Such multimodality is expected to enhance the diversity of sources of innovation (Mariani & Dwivedi, 2024) and foster dynamic, mutually reinforcing creative cycles (Benbya et al., 2024; Eapen et al., 2023). Our studied projects showcase the facilitator's experimentation with multimodality through his use of DALL-E for (imperfectly) translating J.A.K.E.'s concept into a visual format (project #1), helping the team visualize their ideas and drafting the final prototype (project #3). Thus, we encourage practitioners to further explore the potential of multimodal GenAI's capabilities in support of design sprint activities within the solution space. As GenAI rapidly evolves, with new applications emerging continually, we anticipate that maintaining an up-to-date understanding of these advancements will be crucial for maximizing their use in human-centered, collaborative problem-solving processes.

**Navigating the Legal, Ethical, and Technological Landscape.** GenAI tools rely on large language models trained on vast corpora of text, images, and videos. Although they offer powerful 'reservoirs of knowledge' for practitioners, they involve legal, ethical, technological, and privacy concerns. For instance, these models often use data scraped from the internet, sometimes without authorization. Some models may be trained on copyrighted material, raising intellectual property issues (Appel et al., 2023). To mitigate these risks, practitioners could consider using GenAI models that comply with copyright laws or companies that ensure responsible training data sources. Unfortunately, the availability of such models is limited, but data licensing industrial alliances[4] are emerging, third-party auditors[5] are evaluating the provenance of data used for training (Bommasani et al., 2024; Longpre et al., 2024), and datasets designed to minimize copyright and safety concerns are beginning to appear (Arnett et al., 2024; Meyer et al., 2024). In the future, these initiatives may not only improve the transparency and responsible use of datasets for AI training but also support informed decision-making by model users. Second, these models may be prone to bias as they do not inherently grasp contextual knowledge or an organization's specific norms and culture during the innovation process (Gotlar & Hutley, 2024; Hovy & Yang, 2021). Practitioners should keep in mind that they are imperfect models of human culture (Colas et al., 2022), mirroring the norms and values prevalent in their training data, which predominantly originate from Western, Educated, Industrialized, Rich, and Democratic societies (Atari et al., 2023). This can skew the generated content even if used for mediating the human idea generation process, team collaboration, and solution construction. Third, GenAI models have technical limitations that may hinder their contribution to the innovation process, such as generating factually incorrect but plausible-sounding sentences, even when using context files to retrieve historical information generated during the design sprint—for instance, GenAI models' reasoning performance declines as input length increases (Levy et al., 2024). Therefore, practitioners must carefully evaluate and match large language models' capabilities to the specific demands of each task to ensure accurate outcomes and critically assess outputs before sharing them with the collective. Finally, the protection of input data is crucial, especially when handling sensitive company information (IBM, 2024). Current state-of-the-art models are proprietary and offered via online services. Although this enhances GenAI accessibility at lower costs, it introduces data privacy and governance risks, including reduced control over the provider's security compliance regarding input data. The alternative option is to deploy open foundation models on-premises to establish security protocols, but this requires in-house technical capabilities and budget.

[4] https://www.thedpa.ai/

## 5.3 Future Research Avenues

Further exploration into the optimal configurations of human-AI collaborative patterns—referred to as "*superminds*" (Malone, 2018) or "*super teams*"—is essential. Our research has demonstrated the versatility of GenAI as a value-adding collaborator in innovation teams, where it has taken on traditional human roles (Shanahan et al., 2023). For instance, GenAI has functioned as a contributor of baseline ideas, a curator organizing and prioritizing those ideas, an editor refining and presenting content, a back-office assistant handling data management tasks, and a sounding board for testing and iterating concepts. Over time, we observed GenAI evolving as a hyperspecialized, yet nonautonomous *counterpart* within the innovation team (Anthony et al., 2023). The integration of specialized GPTs or multimodal GenAI agents, used throughout the innovation process with autonomous triggers for their activation and coordination, seems increasingly feasible (Ghaffary & Metz, 2024) and suggests a new potential for augmented facilitation (Cvetkovic et al., 2023) together with a tendency to assign part of the facilitation work "in a box" (Briggs et al., 2013). However, this potential raises interesting questions for future research: Who sets the goals for these systems? What exactly are these goals? And crucially, how do we ensure control measures are in place to prevent autonomous agents from making decisions independently? Addressing these questions is critical as we continue to integrate more advanced AI capabilities into team dynamics, ensuring that they complement rather than compromise human input and control.

[5] https://www.dataprovenance.org/

## 6. Conclusion

Our research reveals the extensive potential of integrating GenAI into human-centered innovation processes beyond mere idea generation. Over the course of a year, examining three innovation projects, we observed how the use of GenAI can contribute to both divergent and convergent activities—including problem framing, data synthesis, prioritization, and prototyping. Team members have more time and bandwith to concentrate on tasks where human intuition and contextual understanding are crucial, while GenAI takes on more clerical or repetitive work, resulting in a more streamlined innovation process. That being said, in this rapidly evolving field, researchers and practitioners alike must remain vigilant as AI developments are progressing at an unprecedented pace, with often broad and not fully understood implications.

Looking ahead, we anticipate that the adoption of GenAI will reshape how organizations structure innovation teams and allocate resources. As GenAI techniques mature, new possibilities will emerge for analyzing complex data, generating creative design solutions, and fostering deeper collaboration. However, our study suggests that realizing these benefits will demand a craftsman mindset. Facilitators and collaboration engineers will have to carefully define AI's role, invest in data curation and quality, and build trust in the technology through transparent communication and demonstration of GenAI's capabilities. For practitioners, this will also means rethinking team composition, redefining workflows, and developing strategies to integrate AI-driven tools responsibly. For researchers, our findings highlight important avenues of inquiry—from refining models of human–AI collaboration, to examining ethical and bias-related issues, and to developing robust frameworks for measuring AI's impact on innovation outcomes.

# Appendix A: Original Structure of a Design Sprint

The original DS framework is structured around a five-day process (Knapp et al. 2016), allowing teams to quickly move from concept to concrete feedback, significantly speeding up the traditional development timeline (Gryskiewicz et al., 2018).

Day 1 begins with *framing*, with the team coming together to understand the business challenge. This can involve conducting expert interviews and mapping the user journey, establishing a solid foundation for the entire week, and setting the design sprint goal.

On Day 2, the focus shifts to divergent thinking, where team members individually *sketch out solutions*, generating a mass of ideas and solutions.

By Day 3, these ideas and solutions are collectively curated and narrowed down through a structured selection process, and *the most promising solution is selected* through a voting process.

Day 4 is dedicated to transforming the selected solution into a *prototype* that mimics a real-world product or service as closely as possible. This stage is crucial for turning abstract solutions into tangible, testable artifacts.

The sprint concludes on Day 5, where the prototype is *tested* with users to gather feedback and insights, which are then used to evaluate the concept's viability and decide on the next steps.

## Appendix B: Application of the Key Principles of Interpretive Research in our Work

| **Principle** [1] | **Description** [1] | **Application** |
|---|---|---|
| The Fundamental Principle of the Hermeneutic Circle | This principle suggests that all human understanding is achieved by iterating between considering the interdependent meaning of parts and the whole that they form. This principle of human understanding is fundamental to all other principles. | We applied this principle by iterating between the detailed insights gained at the phase level (across projects) as well as the project level (across phases). This allowed us to understand how parts of the process connected with the broader context of the sprints, and ultimately helped us to derive broadly applicable actionable recommendations. |
| The Principle of Contextualization | Requires critical reflection of the social and historical background of the research setting, so that the intended audience can see how the current situation under investigation emerged. | We considered the background of each innovation project as well as the evolving maturity of GenAI to understand the setting of each design sprint. This included examining how the facilitator's adoption of GenAI tools and techniques evolved from early 2023, providing context on the genesis of the practices observed. |
| The Principle of Interaction Between the Researchers and the Subjects | Requires critical reflection on how the research materials (or "data") were socially constructed through the interaction between the researchers and the participants. | Our collaborative methodology was atypical. It involved dynamic interactions between the practitioner, the sprint participants, and the academic researchers. The practitioner actively participated in the sprints and contributed significantly to the construction of the narrative around GenAI's role in innovation. |
| The Principle of Abstraction and Generalization | Requires relating the idiographic details revealed by the data interpretation through the application of principles one and two to theoretical, general concepts that describe the nature of human understanding and social action. | We abstracted findings from specific GenAI use cases in design sprints to broader implications for innovation practices. Our process of abstraction was driven by the goal of formulating actionable guidelines (not theory) that extend beyond the immediate contexts of our study. |
| The Principle of Dialogical Reasoning | Requires sensitivity to possible contradictions between the theoretical preconceptions guiding the research design and actual findings ("the story which the data tell") with subsequent cycles of revision. | Throughout our research, we remained open to findings that contradicted our initial hypotheses about GenAI's efficacy in design sprints. This involved revising our understanding based on empirical data and ensuring that our |

| Principle [1] | Description [1] | Application |
| --- | --- | --- |
| | | conclusions were robust and reflective of actual practices. |
| The Principle of Multiple Interpretation | Requires sensitivity to possible differences in interpretations among the participants as are typically expressed in multiple narratives or stories of the same sequence of events under study. Similar to multiple witness accounts even if all tell it as they saw it. | We acknowledged and incorporated different perspectives from sprint participants in addition to that of the facilitator, recognizing that each brought unique insights into the use and value of GenAI. This enriched our analysis and allowed for a more nuanced understanding of how GenAI influenced the innovation process. |
| The Principle of Suspicion | Requires sensitivity to possible "biases" and systematic "distortions" in the narratives collected. | We critically evaluated the narratives reported by the sprint facilitator, considering his enthusiasm for GenAI, which could bias the interpretation of its effectiveness. This meant that the academic researchers played the role of a critical sounding board and scrutinized the data for overstatements of success or underestimations of challenges. |

(1) The seven principles and their description reported in this table are from Klein and Myers (1999)

## Appendix C: Timeline of our Collective Sensemaking Sessions

| **Session** | **Date** | **Length (minutes)** | **Focal content discussed during the session** |
|---|---|---|---|
| Project #1 - Spring 2023 | | | |
| 1 | **May** 25, 2023 | 52 | Project #1 (sprint activities, GenAI use, lessons) and envisioned pivots for project #2. |
| Project #2 – Summer 2023 | | | |
| 2 | **July** 13, 2023 | 53 | Project #2 (sprint activities, GenAI use, lessons), comparisons with project #1, and envisioned pivots for project #3. |
| 3 | **October** 5, 2023 | 72 | |
| 4 | **October 31**, 2023 | 72 | |
| 5 | **November 8**, 2023 | 32 | |
| Project #3 – Fall 20233 | | | |
| 6 | **April**, 5 2024 | 120 | Project #2 (sprint activities, GenAI use, lessons) and comparisons with project #1 and #2. |
| 7 | **April**, 8 2024 | 75 | The problem framing practice (GenAI vs Team value, challenges). |
| 8 | **April**, 11 2024 | 100 | The problem definition and team alignment practice (GenAI vs Team value, challenges). |
| 9 | **April**, 12 2024 | 60 | The ideation practice (GenAI vs Team value, challenges). |
| 10 | **April**, 18 2024 | 64 | The problem framing and ideation practices again (GenAI vs Team value, challenges). |
| 11 | **April**, 19 2024 | 66 | The solution choice practice (GenAI vs Team value, challenges). |
| 12 | **April**, 25 2024 | 90 | The solution validation practice (GenAI vs Team value, challenges). |

Appendix D: Data Analysis Process Overview

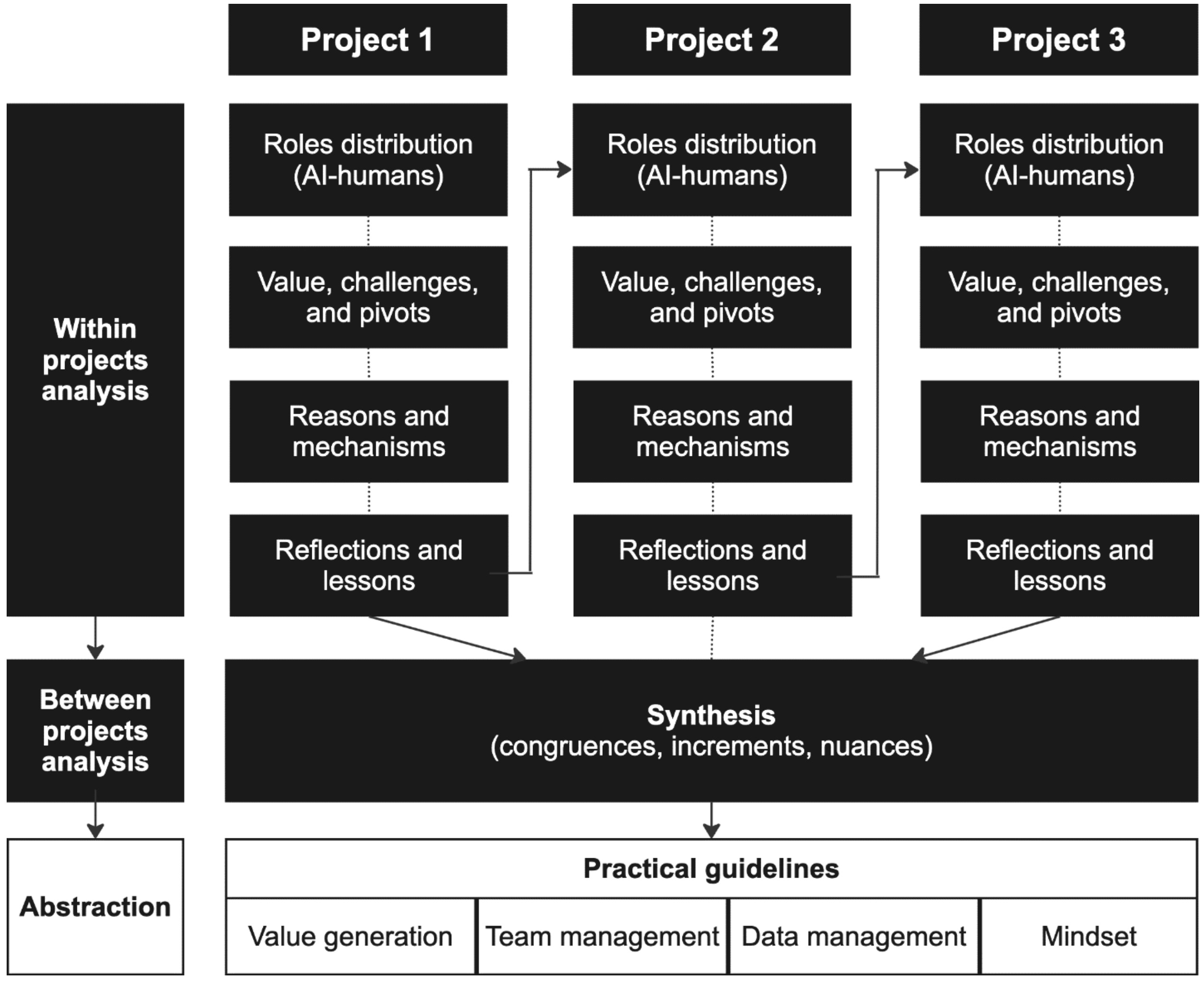